\documentclass[sigconf,nonacm]{acmart}
\AtBeginDocument{%
  \providecommand\BibTeX{{%
    \normalfont B\kern-0.5em{\scshape i\kern-0.25em b}\kern-0.8em\TeX}}}

\copyrightyear{2023}
\acmYear{2023}
\acmDOI{XXXXXXX.XXXXXXX}
\acmConference[SC '23]{The International Conference for High Performance Computing, Networking, Storage, and Analysis}{November 14--19, 2023}{Denver, CO, USA}
\acmBooktitle{SC '23: The International Conference for High Performance Computing, Networking, Storage, and Analysis, November 14--19, 2023, Denver, CO, USA} 
\acmPrice{XX.00}
\acmISBN{978-1-4503-XXXX-X/18/06}

\setcopyright{none}
\usepackage{multirow}
\usepackage{acro}
\usepackage{mathtools}
\usepackage[flushleft]{threeparttable}
\usepackage{tikz}
\usepackage{pgfplots}
\pgfplotsset{compat=newest}
\usepackage{relsize}
\usepackage{colortbl}
\usepackage{siunitx}
\usepackage[justification=centering]{subcaption} 
\usepackage{adjustbox}

\newcommand{\acrodef}[2]{\DeclareAcronym{#1}{short={#1},long={#2}}}
\acrodef{IM}{Ising Model}
\acrodef{KM}{Kuramoto Model}
\acrodef{ODEs}{ordinary differential equations}
\acrodef{POM}{Physical Oscillator Model}
\acrodef{RHS}{right-hand side}
\acrodef{SMT}{simultaneous multithreading}

\usepackage{todonotes}
\definecolor{myblue}{RGB}{37,165,203}
\definecolor{FAUblue}{rgb}{0.000, 0.2196, 0.3961}
\definecolor{myred}{RGB}{175,32,67}

\usepackage{tcolorbox}
\newtcbox{\mybox}{on line,
  colframe=FAUblue,colback=cyan!4!white,
  boxrule=0.5pt,arc=0pt,boxsep=0pt,left=1.6pt,right=1.6pt,top=1.6pt,bottom=1.6pt}

\newcommand{\bytes}{~\si{\byte}~}
\newcommand{\byte}{~\si{\bytes}~}

\newcommand{\second}{~\si{\second}~}

\newcommand{\flop}{~\si{\flop}~}
\newcommand{\flops}{~\si{\flops}~}

\newcommand{\instr}{~\si{\instr}~}
\newcommand{\cycle}{~\si{\cycle}~}
\newcommand{\cycles}{~\si{\cycles}~}

\newcommand{\bit}{~\si{\bit}~}
\newcommand{\bits}{~\si{\bits}~}

\newcommand{\Hz}{~\si{\hertz}~}

\newcommand{\bq}{\begin{equation}}
\newcommand{\eq}{\end{equation}}
\newcommand{\eos}{~.}
\newcommand{\cma}{~,}

\newcommand{\CODE}[1]{\texttt{#1}}  
  
\newcommand{\enquote}[1]{``#1''}

\newcommand{\Heading}[1]{\textbf{#1}}
\newcounter{num}

\usepackage{listings}
\lstdefinestyle{small}{
  language=C++,
  basicstyle=\footnotesize\tt,
  showspaces=false,
  showstringspaces=false,
  breakindent=0pt,
  breaklines,
  captionpos=b,
  breakatwhitespace=true,
  numbers=left,
  numberstyle=\tiny,
  stepnumber=2,
  numbersep=7pt,
  keywordstyle=\color{blue},
  tabsize=2,
}
\lstdefinestyle{smallWithColor}{
  basicstyle=\small\tt,
  language=C++,
  showspaces=false,
  showstringspaces=false,
  breakindent=0pt,
  breaklines,
  captionpos=b,
  breakatwhitespace=true,
  numbers=left,
  numberstyle=\tiny,
  stepnumber=1,
  numbersep=7pt,
  keywordstyle=\color{blue},
  tabsize=2,
  moredelim=**[is][\color{red}]{@}{@},
  moredelim=**[is][\color{FAUblue}]{|}{|},
}

\lstdefinestyle{footnotesizeStyle}{
  basicstyle=\footnotesize\tt,
  language=Matlab,
  showspaces=false,
  showstringspaces=false,
  breakindent=0pt,
  breaklines,
  captionpos=b,
  breakatwhitespace=true,
  numbersep=0pt,
  keywordstyle=\color{blue},
  tabsize=1,
}

\newif\ifblind
\blindfalse

\begin{document}
\title{Physical Oscillator Model for Supercomputing}
\ifblind
\author{Authors omitted for double-blind review process}
\institution{\email{}}
\else
\author{Ayesha Afzal}
\email{ayesha.afzal@fau.de}
\orcid{0000-0001-5061-0438}
\affiliation{%
	\institution{Erlangen National High Performance Computing Center (NHR@FAU)}
	\city{91058 Erlangen}
	\country{Germany}}

\author{Georg Hager}
\email{georg.hager@fau.de}
\orcid{0000-0002-8723-2781}
\affiliation{%
	\institution{Erlangen National High Performance Computing Center (NHR@FAU)}
	\city{91058 Erlangen}
	\country{Germany}}

\author{Gerhard Wellein}
\email{gerhard.wellein@fau.de}
\orcid{0000-0001-7371-3026}
\affiliation{%
	\institution{Department of Computer Science, Friedrich-Alexander-Universität}
	\city{Erlangen-N\"urnberg}
	\country{Germany}
	\postcode{91058}}

\renewcommand{\shortauthors}{Afzal et al.}
\fi
\pagestyle{plain}
\begin{abstract}
A parallel program together with the parallel hardware it is running on is 
not only a vehicle to solve numerical problems, it is also a
complex system with interesting dynamical behavior: resynchronization and
desynchronization of parallel processes, propagating phases of idleness, and
the peculiar effects of noise and system topology are just a few
examples. We propose a physical oscillator model (POM) to describe 
aspects of the dynamics of interacting parallel processes. Motivated
by the well-known Kuramoto Model, a process with its
regular compute-communicate cycles is modeled as an oscillator which is
coupled to other oscillators (processes) via an interaction potential. 
Instead of a simple all-to-all connectivity, we employ a sparse
topology matrix mapping the communication structure 
and thus the inter-process dependencies of the program onto the 
oscillator model and propose two interaction potentials
that are suitable for different scenarios in parallel computing: 
\emph{resource-scalable} and \emph{resource-bottlenecked} 
applications. The former are not limited by a resource bottleneck such
as memory bandwidth or network contention, while the latter are. 
Unlike the original Kuramoto model, which has a periodic
sinusoidal potential that is attractive for small angles,
our characteristic potentials are always attractive for large 
angles and only differ in the short-distance behavior. 
We show that the model with appropriate potentials can mimic
the propagation of delays and the synchronizing and desynchronizing 
behavior of scalable and bottlenecked parallel programs, respectively. 

\end{abstract}

\maketitle

\section{Introduction and related work}

	\subsection{Additive runtime models} 
		
		Analytic white-box (first-principles) performance models are based on a simplified mathematical description of the hardware, the software, and their interactions. They deliver an unmatched level of insight into performance bottlenecks, but they also tend to be imprecise beyond the node level; this is because the intricate interplay between node-level and cluster-level bottlenecks challenges the assumption that the overall runtime is the sum of computation time and communication overhead: $t^\mathrm{total}=t^\mathrm{comp}+t^\mathrm{comm}$~\cite{AfzalHW19}. Many parallel programs do have the regular compute-communicate cycles underlying this model, but they still fail to show additive runtime. 
        The reason is that, irrespective of a perfect translational symmetry, noise, slight load imbalances, hardware topology, and the presence of hardware bottlenecks leads to processes going out of their ``natural'' lock-step~\cite{AfzalHW20,AfzalHW:2022:1}, which may have surprising consequences for performance in applications~\cite{AfzalFGCS23,AfzalHW:2022:4}.

\subsection{Coupled oscillators}\label{sec:motivation}

    The motivation for describing a parallel program in terms of a dynamical
    system emerges from the observation of analogies between the dynamics of coupled
    oscillators and the behavior of processes in bulk-synchronous MPI-parallel codes.
    For example, if a singular effect causes one MPI process to be delayed, 
    the communication dependencies cause this delay to ``ripple'' through 
    the system, a phenomenon called an \emph{idle wave}. Idle waves get damped
    as they travel and will run out eventually~\cite{markidis2015idle}. 
    For parallel programs that
    are not subject to bottlenecks like memory or network bandwidth, this
    behavior leads to a \emph{resynchronization} of the processes~\cite{AfzalHW19}, similar to a swarm of fireflies flashing randomly at first but later
    going into sync~\cite{STROGATZ2000}. Memory-bound programs,
    on the other hand, are prone to \emph{desynchronization} and can develop
    a permanent out-of-sync state that we call a \emph{computational wavefront}~\cite{AfzalHW20}.
    
    A plethora of research has been conducted in engineering and physics about the general phenomenon of self-synchronization among coupled ``oscillating 
    entities.''
    Motivational examples of oscillators are the iconic swarms of fireflies, lasers, the Millennium Bridge, coupled metronomes, neurons, and heart cells. There is an abundance of literature on the paradigmatic Kuramoto model~\cite{kuramoto1975self},
    which is the most popular among the many models proposed for a description of synchronization~\cite{pikovsky2001}.
    Numerous reviews~\cite{STROGATZ2000,arenas2008synchronization,rodrigues2016kuramoto} have been dedicated to the analysis of the Kuramoto model in exploring the onset of synchronization for networks of coupled oscillators.
    This paper uses the Kuramoto Model as a motivation to describe the local and global dynamics of communicating processes in parallel programs on high performance computing clusters.
    
    To the best of our knowledge, describing the execution of a parallel program by mapping it to a collection of coupled harmonic oscillators is a novel idea. The model presented here
    seems to provide a good representation of the qualitative behavior of parallel code, especially in terms of bottleneck behavior, de- and resynchronization, and the propagation of noise. The coupling among the oscillators is implemented as a custom \emph{interaction potential}, which differs from the one in the Kuramoto model, and a \emph{topology matrix}, which describes dependencies between processes and mimics the communication topology of a parallel program. Note that there is an obvious mismatch between the physical concept
    of a harmonic oscillator and an MPI process which ``oscillates'' between
    different states of execution and/or communication. However, this is also
    true for many physical systems for which Kuramoto-like models were used
    to describe collective phenomena. 

    \Heading{Motivation} The physical model is an unknown analogy in the parallel computing field that enables simple and cheap experimentation by solving a system of ODEs rather than running highly parallel programs.
    The model, which we provide as a convenient Matlab application, is fully parameterized.
    Beyond the additive runtime model, a physical model seems more appealing for some scenarios  since it incorporates communication topology, noise, and bottlenecks for the description of parallel distributed computing.

    \Heading{Contributions} In this paper, we show that a collection of coupled
    harmonic oscillators with a suitably chosen interaction potential and 
    connectivity (or \emph{topology}) matrix can reproduce dynamical effects 
    in scalable and memory-bound parallel programs such as propagation and
    decay of idle waves, re- and desynchronization, and the formation of
    computational wavefronts. We also provide a MATLAB implementation\footnote{\url{https://github.com/RRZE-HPC/OSC-AD}\label{foot:AD}}~\cite{POMAD2023} of the
     model that can be used to explore its parameter space.

	\Heading{Overview}
	This paper is organized as follows:
	After covering the motivation for having a physical model for supercomputers, 
	Sec.~\ref{sec:modelsearch} provides details about the search for suitable physical analogs.
	Sec.~\ref{sec:POM} covers the description of the physical model and the analogy between the model and the parallel programs.
	After providing the details about
	our experimental environment and methodology (Sec.~\ref{sec:environment}),
	the evaluation is done in Sec.~\ref{sec:evaluation}.
	Finally, Sec.~\ref{sec:conclusion} discusses the findings and provides an outlook to future work.

\section{Search for possible physical models}\label{sec:modelsearch}
	Many analogies are thinkable between interacting processes in parallel programs and physical models.
	Here, we briefly discuss the possible options, highlighting their strengths and weaknesses.
	
\subsection{Continuum models}
		There are a lot of possible continuum models which include wave equations with dissipative and non-linear parts~\cite{HUANG20067014,HUNTER1988253} and which may serve as blueprints of wave-like behavior in parallel programs. For instance, solitons~\cite{drazin1989solitons} are solutions of weakly nonlinear dispersive PDEs. They are stabilized by a cancellation of nonlinear and dispersive effects in the medium and propagate at a constant velocity and are similar in character to idle waves.
		Other motivational examples of PDEs include Navier-Stokes and shallow-water equations, etc.~\cite{potter2021schaum,bresch2009shallow}.
		Despite the appeal of such analogies, it is as yet unclear how a continuum limit of the discrete MPI processes in parallel programs should be performed.

\subsection{Discrete models}
		We considered the following two discrete models to look for a physical analogy to the dynamics of parallel programs.
		
        \subsubsection{\ac{IM}}
		Models like the Ising model~\cite{ising1924beitrag} describe interacting entities (spins in the Ising case) and show interesting effects like symmetry breaking, phase transitions, and long-range correlations.
		The system tends to assume a state of lowest energy. Heat disturbs this tendency, thus creating the possibility of different structural phases.
		Since this model lacks the necessary natural oscillating behavior (processes alternate at least between computation and communication), it seems unsuitable for the desired analogy. 
		
        \subsubsection{\ac{KM}}
		The Kuramoto model~\cite{kuramoto1975self,kuramoto1984chemical} is an iconic, well-studied model of self-organizing behavior on which a lot of literature exists, including its continuum limit.
        It describes a collection of $N$ oscillators with natural frequencies $\{\omega_i\}$ which are coupled via a periodic interaction potential that depends on the mutual phase differences $(\theta_j(t)-\theta_i(t))$:
		\bq\label{eq:KM}
		\dot{\theta_i}(t) = \omega_i + \frac{K}{N} \sum_{j=1}^{N} sin (\theta_j(t)-\theta_i(t)),\quad\forall{i = 1 \cdots N}
		\eq

        The coupling strength is globally parameterized by the parameter $K$.
		There are several reasons why the plain Kuramoto Model is unsuitable for describing parallel programs.
  
		First, it implements full connectivity, i.e., a global, all-to-all coupling of the oscillators, which is an undesirable pattern in parallel computing. In fact, a parallel program with this pattern shows extremely fast idle wave propagation and no potential for natural process desynchronization because the all-to-all connectivity acts like a synchronizing barrier in each time step (analogous to one oscillator period). 
		Later research modified the model by incorporating varying coupling strengths between pairs of oscillators, which can be seen as a kind of network topology~\cite{rodrigues2016kuramoto, Tommaso2020}. We will come back to this idea later.

		Second, the periodic sinusoidal interaction potential in the \ac{KM} leads to the onset of self-synchronization of the oscillators, a phenomenon that occurs also in resource-scalable parallel pro\-grams~\cite{AfzalHW20}.\footnote{Processes are resource scalable if either there are no bottlenecks (such as memory bandwidth) or the execution of the parallel program does not expose them.}
		However, resource-bottlenecked (e.g., memory-bound, communi\-cation-bound) parallel programs also exhibit spontaneous-onset desynchronization~\cite{AfzalHW20}, which the original Kuramoto Model does not show.

        Third, the periodicity of the KM potential allows for \emph{phase slips}, i.e., two oscillator phases can be a multiple of $2\pi$ apart without changing the dynamics. This is impossible with periodically communicating processes where usually a computation phase cannot start before a message has been received. 
    	Finally, parallel computing systems are subject to many sources of noise, which the plain KM does not know. However, later research modified the model to include it~\cite{rodrigues2016kuramoto}.
\section{Physical Oscillator Model (POM)}\label{sec:POM}
    \subsection{Theoretical description}
    We use the following modified coupled oscillator model to describe $N$ parallel MPI processes on a cluster:
    	\bq\label{eq:POM}
    	\dot{\theta_i}(t) = \frac{2\pi}{t^\mathrm{comp} + t^\mathrm{comm}}
    	+ \zeta_i (t)
    	+ \frac{v_p}{N} \cdot \sum_{j=1}^{N} T_{ij} V \left(\theta(t, \tau_{ij}(t))\right)
    	\eq   
 
    The intrinsic angular frequency $\omega_i$ of the oscillators in the plain Kuramoto Model is substituted by the frequency of the processes alternating between computation and communication phases of duration $t^\mathrm{comp}$ and $t^\mathrm{comm}$, respectively. 
	\emph{Noise} in MPI can take different forms; here we include time-dependent process-local noise with some distribution $\zeta_i(t)$ and interaction noise $\tau_{ij}(t)$, i.e., random delays caused by varying communication time. The former is implemented as a jitter in the local oscillator frequency and can also serve to model load imbalance, while the latter impacts the phase difference $\theta(t, \tau_{ij}(t)) = \theta_j(t-\tau_{ij}(t)) - \theta_i(t)$~\cite{rodrigues2016kuramoto}.
    The coupling strength $v_p=\frac{\beta\cdot\kappa}{t^\mathrm{comp} + t^\mathrm{comm}}$ is motivated by the connection between idle wave speed and communication characteristics as described in ~\cite{AfzalHW2021}:
	Messages sent via the eager (rendezvous) protocol have $\beta=1~(2)$, and $\kappa$ is the sum over all communication distances.
	However, if the outstanding non-blocking MPI requests of all communication partners are grouped in the same \CODE{MPI\_Waitall}, the parameter $\kappa$ becomes equal to longest distance only~\cite{AfzalHW2021}.
    The topology matrix $T_{ij}$ together with the interaction potential $V$ describes how processes interact: We have $T_{ij} = 1$ if there is a connection between oscillators $i$ and $j$, and $T_{ij} = 0$ otherwise (see Fig.~\ref{fig:MPIprogram}). The shape of the potential $V(\cdot)$ can be used to distinguish between synchronizing (i.e., Kuramoto-like) and desynchronizing behavior.
	
\subsection{Model implementation}
	To solve the first-order coupled ODEs system (\ref{eq:POM}) we use a robust explicit Runge-Kutta (4,5) method (Dormand-Prince). We provide a full implementation of the solver, including a graphical interface for easy experimentation, as a Matlab code in the artifact appendix\footref{foot:AD}.

	It is a flexible tool to study the influence of the model parameters and allows to set different initial conditions (synchronized, desynchronized), noise (both variants in our model), one-off delays, coupling strengths, and communication topology.  
    
	Three options for result visualization are provided: (i) the \emph{circle diagram}, where colors represent the different frequencies, with blue being fast and yellow being slow, (ii) the \emph{timeline of phase differences} for oscillators, and (iii) the \emph{timeline of potentials} (see Listings in the artifact appendix\footref{foot:AD}).
	To demonstrate analogy between MPI program and model, in the standard view we choose the \emph{timeline of oscillator phases $\theta_i-\omega t$} normalized to the slowest (``lagger'') process as the baseline.

\section{Test bed and Experimental Setup} \label{sec:environment}
We use MPI-parallel toy codes for comparisons with the oscillator model. They include MPI point-to-point communication via \CODE{MPI\_Irecv}, \CODE{MPI\_Send}, and \CODE{MPI\_Wait*}.
    To mimic \emph{resource-scalable parallel programs}, we use a ``PISOLVER'' code which calculates the value of $\pi$ by computing $\int_0^14/(1+x^2)\,\mathrm dx$ using the midpoint rule and \SI{500}{\mega~steps}.
    For \emph{resource-bottlenecked parallel programs} 
	we run pure-MPI versions of the STREAM Triad~\cite{mccalpin1995memory}
    \texttt{A(:) = B(:) + s*C(:)}
    and of the \enquote{slow} Sch\"onauer Triad
    \CODE{A(:) = B(:) + {cos(C(:)/D(:))}}, for which the low-throughput cosine and floating-point division shifts the bandwidth saturation point to a higher number of cores. The scaling behavior across the cores of one socket is shown in Fig.~\ref{fig:potential_scalability}(b) for the three codes.
    In all cases we add MPI with short messages after each sweep to implement different communication topologies. The benchmark results in terms of delay propagation, de-, and resynchronization were published in previous work~\cite{AfzalHW19, AfzalHW20, AfzalHW2021}. We present the results on one of the benchmark systems ({Meggie}\footnote{\ifblind{URL redacted for double-blind review.}\else\url{https://hpc.fau.de/systems-services/documentation-instructions/clusters/meggie-cluster}\fi}) here for reference; the results for the other system (SuperMUC-NG) can be found in the artifact appendix\footref{foot:AD}.
        \enquote{Meggie} is a fat-tree \SI{100}{\giga \bit \per \second} Omni-Path cluster with dual-socket nodes (\SI{64}{\giga \byte} DDR4) comprising ten-core Intel Xeon
		``Broadwell'' E5-2630v4 CPUs (\SI{2.2}{\giga \Hz}, \SI{68}{\giga \byte / \second}).
    We ran programs with $40$ and $18$ MPI processes on $4$ and $2$ sockets of the Meggie system, respectively.
    Working sets for the memory-bound codes were chosen large enough to not fit into any cache, i.e., at least 10$\times$ the last-level cache size.
    We refer to artifact appendix\footref{foot:AD} for further details on the experimental setup, the methodology, and the hardware and software environment.

\section{Evaluation and Implications} \label{sec:evaluation}
	To evaluate the modeling power of our novel coupled oscillator model, we describe its connections to MPI program phenomenology. Figure~\ref{fig:MPIprogram} visualizes some of the corner cases with next-neighbor (top row) and next-plus-next-next neighbor (bottom row) communication and scalable (left column) versus saturating (right column) codes, using 
    MPI traces (inner images) and phase visualizations (circles).

 	\begin{figure}[tb]
	\centering
 	\hspace{-0.7em}
	\begin{minipage}[c]{0.25\textwidth}
        \begin{tikzpicture}
\begin{axis}[
    trim axis left, trim axis right, scale only axis,
	axis background/.style={fill=white!97!black},
	width=0.96\textwidth,height=0.68\textwidth,
    xmin=-10, xmax=10,
    ymin=-1.2, ymax=1.99,
    axis lines=center,
    domain=-10:10,
    ylabel={Potential$, V(\theta_j-\theta_i)$},
    xlabel style={align=center},
    xlabel= {Phase difference$,$\\ $\theta_j-\theta_i$},
    	ymajorgrids,
		tick style={thin},
		x label style={at={(1.02,0.02)},font=\scriptsize},
		y label style={font=\scriptsize},
		x tick label style={font=\scriptsize},
		y tick label style={xshift={(\tick==1)*0.4em},font=\scriptsize},
		y tick label style={xshift={(\tick==-1)*0.4em},font=\scriptsize},
		legend columns = 1, 
		legend style = {
			nodes={inner sep=0.117em},
			draw=none,
			fill=none,
			font=\scriptsize,
			cells={align=left},
			anchor=east,
			at={(0.5,0.85)},
			/tikz/column 1/.style={column sep=5pt,},
		},
    ]
    \addplot+ [mark=none,draw=red, thick,samples=100] {tanh(x)};
    \addlegendentry{{Scalable}}
    \addplot+ [mark=none, draw=blue, thick, samples=100]
    {(-sin(deg(x))*(x<4.7)*(x>-4.7)))
    +((x>4.7)-(x<-4.7)))};
    \addlegendentry{{Bottlenecked}}
    \node [right,  thick, blue, font=\bfseries] at (axis cs: 4.3,1.2) { $\sigma$};
\end{axis}
\end{tikzpicture}
		\caption*{\footnotesize (a) Potential visualization}
	\end{minipage}
	\hspace{5.5pt}
	\begin{minipage}[c]{0.22\textwidth}
	\begin{tikzpicture}
		\begin{axis}[trim axis left, trim axis right, scale only axis,
		axis background/.style={fill=white!97!black},
		width=0.72\textwidth,height=0.5\textwidth,
		xlabel = {{{Processes per Meggie socket}}},
		ylabel = {{{Memory bandwidth [\si{\mega \byte / \second}]}}},
		xmin=0,
		xmax=10,
		ymin=0,
		ymax=99000,
		y label style={at={(-0.1,0.4)},font=\scriptsize},
		x label style={font=\scriptsize},
		x tick label style={font=\normalsize},
		y tick label style={font=\normalsize},
		ymajorgrids,
		tick style={thin},
		xtick pos=left,
		xtick={0,2,4,6,8,10},
		ytick={0,20000,40000,60000},
		legend columns = 1, 
		legend style = {
			nodes={inner sep=0.117em},
			draw=none,
			fill=none,
			font=\tiny,
			cells={align=left},
			anchor=east,
			at={(0.96,0.79)},
			/tikz/column 1/.style={column sep=5pt,},
		},
		set layers, 
		]
        
        \begin{pgfonlayer}{axis background}
			\fill[shade, left color=FAUblue!6, right color=FAUblue!6]
			(rel axis cs:0,0)--(rel axis cs:0.5,0)--
			(rel axis cs:0.5,0.6)--(rel axis cs:0,0.6)--cycle;
			\fill[shade, left color=FAUblue!14, right color=FAUblue!14]
			(rel axis cs:0.5,0)--(rel axis cs:0.9,0)--
			(rel axis cs:0.9,0.6)--(rel axis cs:0.5,0.6)--cycle;
			\fill[shade, right color=FAUblue!36, left color=FAUblue!36]
			(rel axis cs:0.9,0)--(rel axis cs:1,0)--
			(rel axis cs:1,0.6)--(rel axis cs:0.9,0.6)--cycle;
		\end{pgfonlayer} 
		
		\addplot+[mark size=1pt,error bars/.cd, y dir=both, y explicit]
		table
		[
		x expr=\thisrow{Cores}, 
		y error minus expr=\thisrow{Median}-\thisrow{Min},
		y error plus expr=\thisrow{Max}-\thisrow{Median},
		row sep=crcr]{
		Cores	Median		Min		Max\\
1	16081.349	16081.349	16081.349\\
2	28282.9129	28282.9129	28282.9129\\
3	38011.239	38011.239	38011.239\\
4	45320.711	45320.711	45320.711\\
5	49324.4899	49324.4899	49324.4899\\
6	51933.2305	51933.2305	51933.2305\\
7	52217.5162	52217.5162	52217.5162\\
8	53376.9859	53376.9859	53376.9859\\
9	52863.7581	52863.7581	52863.7581\\
10	53463.242	53463.242	53463.242\\
		};
		\addlegendentry{{STREAM}}

		
		\addplot+[myblue,mark options={fill=myblue},mark size=1pt,error bars/.cd, y dir=both, y explicit,]
		table
		[
		x expr=\thisrow{Cores}, 
		y error minus expr=\thisrow{Median}-\thisrow{Min},
		y error plus expr=\thisrow{Max}-\thisrow{Median},
		row sep=crcr
		]{
		Cores	Median		Min		Max\\
1	8296.5011	8296.5011	8296.5011\\
2	15875.1178	15875.1178	15875.1178\\
3	22530.8013	22530.8013	22530.8013\\
4	29022.0039	29022.0039	29022.0039\\
5	34677.882	34677.882	34677.882\\
6	40004.6952	40004.6952	40004.6952\\
7	44520.1861	44520.1861	44520.1861\\
8	48264.4511	48264.4511	48264.4511\\
9	51905.5896	51905.5896	51905.5896\\
10	52738.2617	52738.2617	52738.2617\\
        };
		\addlegendentry{{Slow Sch\"onauer}}
		
		\addplot+[red, no marks,error bars/.cd, y dir=both, y explicit,]
		table
		[
		x expr=\thisrow{Cores}, 
		y error minus expr=\thisrow{Median}-\thisrow{Min},
		y error plus expr=\thisrow{Max}-\thisrow{Median},
		row sep=crcr
		]{
		Cores	Median		Min		Max\\
1	4296.5011	4296.5011	4296.5011\\
2	8593.0022	8593.0022	8593.0022\\
3	12889.5033	12889.5033	12889.5033\\
4	17186.0044	17186.0044	17186.0044\\
5	21482.5055	21482.5055	21482.5055\\
6	25779.0066	25779.0066	25779.0066\\
7	30075.5077	30075.5077	30075.5077\\
8	34372.0088	34372.0088	34372.0088\\
9	38668.5099	38668.5099	38668.5099\\
10	42965.011	42965.011	42965.011\\
        };
		\addlegendentry{{PISOLVER}} 
 		\end{axis}
 	\end{tikzpicture}
 	\caption*{\footnotesize (b) Scalability}
	\end{minipage}
	\caption{(a) Potential $V((\theta_j-\theta_i))$ for scalable (red) and non-scalable (blue) parallel programs. 
    In the latter, the position of the first zero defines the stable desync state. (b) Scalability of the MPI-parallel micro-benchmarks on Meggie cluster.}
    \label{fig:potential_scalability}
\end{figure}

    
	\subsection{Delay propagation} 
    ~\par

    \subsubsection{Resource-scalable parallel programs}
    Figure~\ref{fig:MPIprogram}  shows MPI traces obtained using Intel Trace Analyzer and Collector (ITAC) with computation (white) and communication (red) Within the circular phase visualizations of the physical oscillator model. Idle waves originating from a one-off delay (extra workload performed by the 5th MPI process in blue) ripple through the parallel program due to the dependencies between processes mediated via MPI communication.
	This is called \emph{delay propagation}. Idle waves interact nonlinearly with each other and with system noise, leading to their eventual decay.
    Communication properties (periodicity, range), system noise, and system topology all have a huge impact on the speed and general behavior of the wave~\cite{AfzalHW2021}, but the most important property is its speed. Very weak or nonexistent coupling (corresponding to $\beta\kappa\approx 0$ in the model (Eq. \ref{eq:POM})) corresponds to free processes with no dependencies. The case $\beta\kappa=1$ describes next-neighbor coupling and minimum idle wave speed, leading to slow relaxation into a synchronized state. The larger $\beta\kappa$ the faster the wave and the ``stiffer'' the system gets, until with very large $\beta\kappa$ the system is strongly synchronizing since any variation is immediately propagated through the whole system.

    \subsubsection{Resource-bottlenecked parallel programs}
    The propagation speed of idle waves is influenced by the coupling in the same way for non-scalable as for scalable programs. One major difference is that for memory-bound code, idle waves have an additional decay mechanism even under noise-free conditions, and after the idle wave has run out, a residual computational wave with desynchronized execution remains; see Fig.~\ref{fig:MPIprogram}(b, d).

	\subsection{Connecting scalability and model potential}
 ~\par
    
    \subsubsection{Self-synchronization in resource-scalable parallel code} 
\begin{figure}[tb]
    \centering
    \includegraphics[scale=0.35]{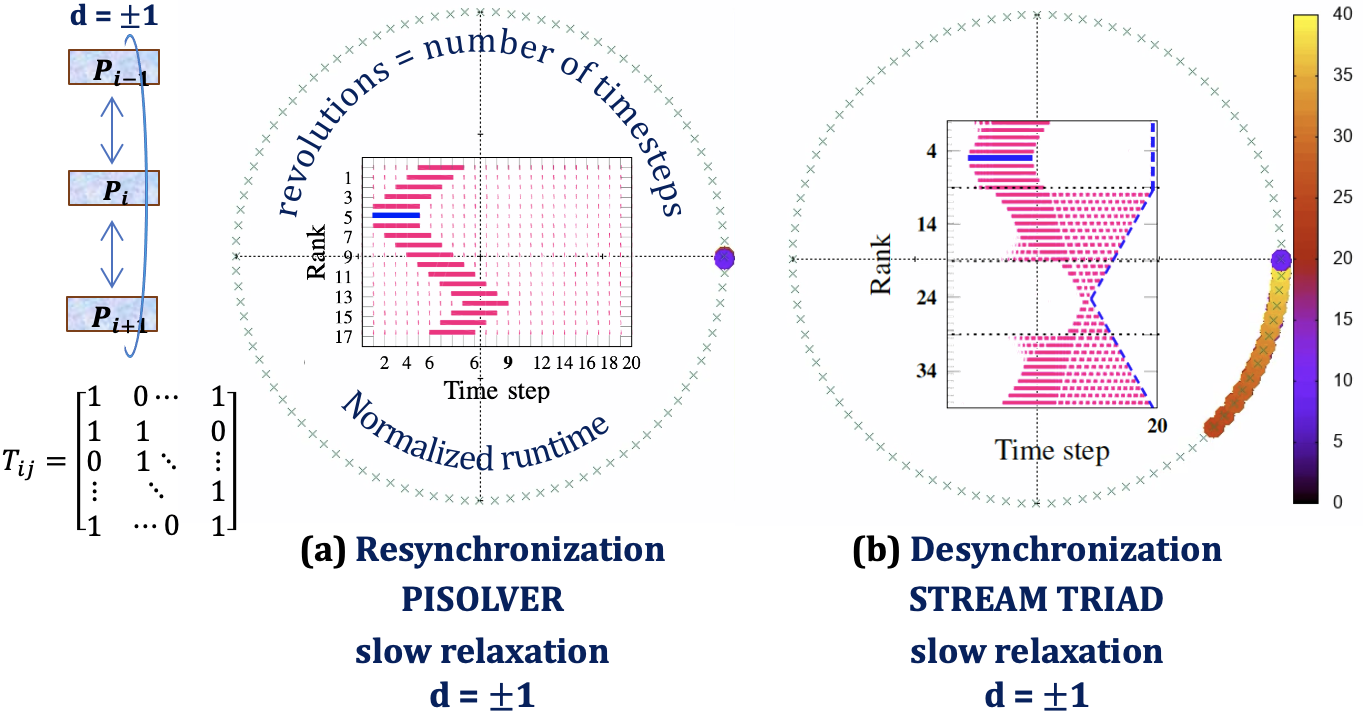}
    \includegraphics[scale=0.35]{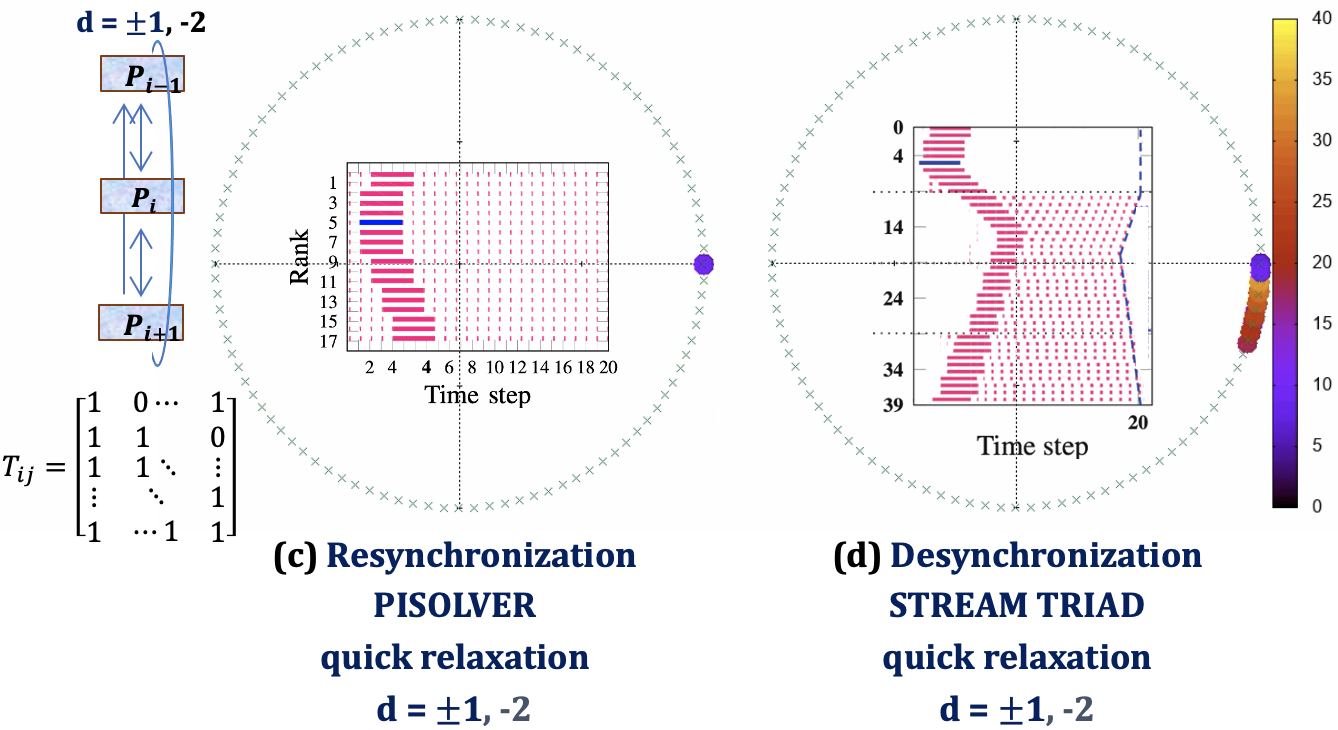}
    \caption{Analogy between MPI programs and physical oscillator model for the communication typologies of $d=\pm 1$ (top) and $d=\pm 1,-2$ (bottom). Inside the circles we show MPI traces of idle waves progressing. The dots on the circle show the asymptotic state of the phase differences among the oscillators in the model. See text for more details. Video animations are available at \url{http://tiny.cc/MPI_triad}.}
    \label{fig:MPIprogram}
\end{figure}

    Bulk-syn\-chro\-nous, re\-source-scalable applications exhibit self-re\-synchro\-ni\-zation on an undisturbed system, i.e., the coupling favors a close proximity of the phases.
    A disturbance or ``pull'' causes phase differences across oscillators, but the system ``snaps back'' into a synchronized state, i.e., all oscillators are eventually in phase with the same natural frequency $\omega_i$; see Fig.~\ref{fig:MPIprogram}(a, c).
    The interaction potential in the oscillator model should be able to mimic this behavior, but
	the periodic sinusoidal potential of the plain \ac{KM} is unsuitable since it allows for phase slips (see above) and has zeros at multiples of $\pi$. One option is a hyperbolic tangent; red line Fig.~\ref{fig:potential_scalability}(a): 
    \bq\label{eq:attractive}
    V (\theta_j-\theta_i) = \tanh\left(\theta_j(t) - \theta_i(t)\right)\eos
    \eq
    
	This potential forces oscillators with any phase difference into sync, which is the property we are looking for.

    \subsubsection{Self-desynchronization in resource-bottlenecked parallel code}		
    Bulk-synchronous bottlenecked programs on a silent system avoid the bottleneck by drifting out of lockstep.
    In other words, the translationally symmetric state is unstable and any slight disturbance blows up and leads to a broken-symmetry state~\cite{AfzalHW20}.
    In the model, this means that each oscillator still has the same frequency but there is a phase difference between adjacent oscillators; see Figs.~\ref{fig:MPIprogram}(b, d) for the asymptotic state of the phase differences among the oscillators in the circular phase diagram and the runtime differences among processes on three of four Meggie sockets in the MPI traces. 

    There are many interaction potentials that would show this property; the requirement is to have attraction at long range but repulsion at short range. We use the following function:
    \bq\label{eq:repulsive}
    V(\theta_i-\theta_j)
	= \left\{\begin{array}{lc}
      -\sin\left(\frac{3\pi}{2\sigma}\cdot(\theta_i-\theta_j)\right) & \mbox{if} \left|\theta_i-\theta_j\right|<\sigma \\
      \mbox{sgn}\left(\theta_i-\theta_j\right) & \mbox{else}
    \end{array}\right.\cma
    \eq
    where $\sigma$ acts as an ``interaction horizon'' which marks the transition to a constant potential in either direction; see the blue line in Fig.~\ref{fig:potential_scalability}(a). A small $\sigma$ corresponds to almost synchronized oscillators, or a ``stiff'' parallel code with long-range, synchronizing communication since the phase differences settle at the first zero of the potential, which is at $2\sigma/3$. Large $\sigma$ describes strong desynchronization with short-range dependencies.  The ``interaction horizon'' $\sigma$ is thus directly correlated with the idle wave propagation speed and the phase spread.
    For instance, increasing the communication stiffness from Figure~\ref{fig:MPIprogram}(b) to Figure~\ref{fig:MPIprogram}(d) led to a threefold increase in the speed of delay propagation (seen in the MPI traces) and a corresponding decrease in oscillator phase spread in the asymptotic state (circular phase diagrams).

\section{Discussion and future work} \label{sec:conclusion}
Motivated by the well-known Kuramoto model, we have suggested a description of MPI-parallel, bulk-synchronous and barrier-free codes using a novel  physical model comprising coupled oscillators. The coupling is described by a topology matrix and an interaction potential, where the former mimics the dependencies between processes enforced by MPI communication and the latter is used to model different scaling behavior on the system's inherent hardware bottlenecks. We introduced potentials for those two cases which can reproduce the asymptotic behavior of memory-bound and scalable programs under one-off noise injections and subsequent idle wave propagation: resynchronization, i.e., re-establishing the translationally symmetric state, and desynchronization, i.e., settling in a state of broken translational symmetry. The potentials were designed so that different communication behavior in the MPI program can be described in a qualitative way. 

As the parallelism of HPC clusters grows, the inherent node-level and network-level bottlenecks become more difficult to assess, and their interaction with parallel code causes counter-intuitive effects. Furthermore, frequently synchronizing parallel programs are incompatible with massive parallelism; in the future, parallel code may be more strongly task-based and asynchronous, allowing for slow idle wave progression and desynchronization. Since the number of model parameters is very small, the physical oscillator model provides a simple way to characterize systems where collective behavior is relevant for program performance. If a well-defined continuum limit of the model can be found, it could be useful in hardware-software co-design for expensive components like network infrastructure.\footnote{This idea goes back to Richard Feynman and his analysis of the network routers in the Connection Machine: \url{https://longnow.org/essays/richard-feynman-connection-machine}} The deviation from lock-step behavior could help to evade bottlenecks, allowing for cheaper design points in the hardware.

Although the idea of mapping a parallel program to a physical oscillator model is novel and appealing, it is as yet unclear whether more refined models and/or potentials could be found that do a better job at mimicking parallel program dynamics; for instance, we have not yet explored the role of the noise functions that the physical model includes and whether these would be able to properly describe idle wave decay. It is also unknown whether the symmetry-breaking transition observed for non-scalable programs (and in the model) is connected to a \emph{Goldstone mode}. These questions are subject to future work.

\ifblind
\else
\begin{acks}
The authors gratefully acknowledge the scientific support and HPC resources provided by the Erlangen National High Performance Computing Center (NHR@FAU) of the Friedrich-Alexander-Universität Erlangen-Nürnberg (FAU) and LRZ Gar\-ching. NHR funding is provided by federal and Bavarian state authorities. The hardware at NHR@FAU is partially funded by the German Research Foundation (DFG) -- 440719683. This work was partly supported by the Competence Network for Scientific High-Performance Computing in Bavaria (KONWIHR) under project ``OMI4Papps.''
\end{acks}
\fi

\bibliographystyle{ACM-Reference-Format}
\bibliography{references}

\end{document}